# Dynamical spin injection at a quasi-one-dimensional ferromagnet-graphene interface

S. Singh[1], A. Ahmadi[1], C.T. Cherian[2,4], E. R. Mucciolo[1], E. del Barco[1], and B. Özyilmaz[2,3,4,5]

[1]Department of Physics, University of Central Florida, Orlando, Florida USA, 32816

[2]Department of Physics, National University of Singapore, 2 Science Drive 3, Singapore 117542

[3]NanoCore, 4 Engineering Drive 3, National University of Singapore, Singapore 117576

[4]Graphene Research Center, National University of Singapore, Singapore 117542

[5]NUS Graduate School for Integrative Sciences and Engineering (NGS), National University of Singapore, Singapore 117456

**Abstract:** We present a study of dynamical spin injection from a three-dimensional ferromagnet into two-dimensional single-layer graphene. Comparative ferromagnetic resonance (FMR) studies of ferromagnet/graphene strips buried underneath the central line of a coplanar waveguide show that the FMR linewidth broadening is the largest when the graphene layer protrudes laterally away from the ferromagnetic strip, indicating that the spin current is injected into the graphene areas away from the area directly underneath the ferromagnet being excited. Our results confirm that the observed damping is indeed a signature of dynamical spin injection, wherein a pure spin current is pumped into the single-layer graphene from the precessing magnetization of the ferromagnet. The observed spin pumping efficiency is difficult to reconcile with the expected backflow of spins according to the standard spin pumping theory and the characteristics of graphene, and constitutes an enigma for spin pumping in two-dimensional structures.

The efficient generation of pure spin currents holds a great deal of promise for spintronics applications, with several existing methods already demonstrated, such as the spin Hall effect [1], electrical spin injection [2], voltage-based spin pumping[3] dynamical spin pumping [4,5], and optical generation of spin packets[6]. Particularly, the dynamical generation of spin currents carries special interest because no net charge current is involved in the process. In this method, spin angular momentum is transferred from the precessional magnetization in a ferromagnet (FM) to an adjacent non-magnetic (NM) system. This approach has already been experimentally demonstrated in several FM/NM interfaces, including NM systems such as metals[7], semiconductors [5,8] or organic based materials[9]. A few advantages of this method are that it does not suffer from the impedance mismatch problem, it is scalable to large samples, it provides a high spin injection efficiency, and it is not based on the spin-orbit coupling to function, allowing its employment in systems without this interaction.

Within the exciting current trend to explore novel low-dimensional systems, the possibility to inject pure spin currents in graphene and other two-dimensional (2D) crystals has attracted considerable attention in the past few years. In particular, graphene rises as a prototype system to explore this physics due to its crystalline nature, excellent electronic properties[10], tunable spin-orbit coupling (*i.e.*, via adatom engineering)[11,12] and long spin relaxation lengths[13]. In addition, graphene can act as a high-fidelity channel for spin information transfer (due to its small intrinsic spin-orbit coupling and absence of nuclear spins), as well as provide the platform for electrical manipulation of the spin polarization (*i.e.*, on-demand enhancement of spin-orbit coupling[11,12]). Dynamical spin injection in FM/graphene (FM/Gr) interfaces has been recently demonstrated [14,15]. In the original work, we associated the observed enhancement in dynamical damping of extended FM/Gr films to the generation of pure spin current in graphene resulting from losses of spin angular momentum in the ferromagnet (*i.e.*, spin pumping[14]). Subsequently, the dynamical injection of spin currents in graphene was demonstrated by spin-charge conversion measurements in a Pd strip placed laterally and in close proximity to a FM/Gr interface undergoing FMR[15]. Although providing evidence for spin injection, none of these works shed light into the real nature of spin pumping at the FM/Gr interface. Estimates of the spin-mixing conductance obtained from the broadening of the FMR peaks resulted in surprisingly high values (*e.g.*, $g_{\uparrow\downarrow} = 5.26\times10^{19}$ m$^{-2}$ from $\Delta\alpha_{Py/Gr} = \alpha_{Py/Gr} - \alpha_{Py} = 1\times10^{-2}$), comparable to systems with high spin-orbit coupling[7] (such as heavy metals Pt and Pd). In addition, the direct deposition of the ferromagnetic Permalloy (Py) film on top of the graphene layer could cause magneto-structural[16] changes in the Py surface and a subsequent increase in straight fields altering the spin dynamics and accumulation in the semiconductor[17], responsible for the observed change in damping when comparing with films without graphene, *i.e.*, deposited on the bare wafer. A direct measurement of the spin Hall angle in FM/Gr interfaces has not been reported yet. A recent work by R. Ohshima *et al.*[18] published during the preparation of this manuscript, claims the observation of the ISHE signal in single-layer graphene upon conversion of a spin current pumped from an extended insulating ferromagnet (*i.e.*, YIG) into a charge current. According to that report, the spin current is injected perpendicularly to the YIG/Gr interface, and the ISHE electric field measured within the graphene plane, as is the norm in other ferromagnet/conductor heterostructures. However, we find this interpretation rather questionable, since conduction perpendicular to a single-layer graphene in not possible simply due to the lack of a third dimension. Since conduction perpendicular to the graphene sheet is not possible, the geometry proposed by the authors, where the ISHE is measured along the graphene plane, is impracticable.

The observations in that work could be explained in terms of alternative physics, such as an inverse Rashba-Edelstein effect[19] or similar, but never in terms of the ISHE.

In this letter we present FMR experiments performed on different FM/Gr interfaces designed to systematically identify and eliminate damping enhancement arising from processes other than spin pumping. In particular, a substantial enhancement of the Gilbert damping observed in Py/Gr strips when the graphene layer protrudes a few micrometers away from the edges of a narrow Py strip is univocally associated to spin pumping at the quasi-one-dimensional interface between the Py edge and graphene, which shows lower spin-mixing conductance values ($g_{\uparrow\downarrow} = 6.89 \times 10^{18}$ m$^{-2}$) than in extended films but still comparable to those obtained in Py/Pt interfaces (*e.g.*, $g_{\uparrow\downarrow} = \sim 1\text{-}4 \times 10^{19}$ m$^{-2}$). We also provide a theoretical analysis which shows the observed spin injection efficiency to lie well beyond that expected from the spin conduction channels provided by single-layer graphene, opening fundamental questions about the nature of spin injection into this 2D crystal.

The graphene layers used in our experiments are grown by the standard chemical vapor deposition (CVD) method on thin Cu foils[20]. Graphene is subsequently transferred onto the substrate using a wet chemistry process and characterized by Raman spectroscopy. We use 14 nm-thick films of Ni$_{20}$Fe$_{80}$ Permalloy (deposited by e-beam evaporation in high vacuum conditions) as the ferromagnet for all our studies. For FMR on extended films, we place the sample upside-down on the central part of our broad-band μ-CPW FMR sensor, which is coated with an insulating polymer to prevent electrical contact with the sample[14]. For FMR on patterned films in the shape of long and narrow Py/Gr strips, the sample is buried directly underneath the central line of the CPW, isolated from it by a thick (~100 nm) insulating layer of oxide.

Before getting into the detailed discussion of the main results of this work, we want to briefly discuss a set of experiments designed to address the effect of magneto-structural changes in the surface of the Py due to the immediate presence of graphene underneath, which could cause a non-dynamical broadening of the FMR [21]. In the first experiment, the stacking order of the Py and graphene layers has been reversed with respect to the original experiments[14], where the Py was deposited directly on top of the graphene layer (*i.e.*, a Py/Gr stacking). In the present case, the Py film is deposited on a bare Si wafer coated with 300nm of thermally grown $SiO_2$ and graphene transferred on top afterwards (*i.e.*, a Gr/Py stacking), with the objective of maintaining the Py film unaltered by the presence of graphene. A clear enhancement of the Gilbert damping is obtained when graphene is present (*i.e.*, $\Delta\alpha_{Gr/Py} = \alpha_{Gr/Py} - \alpha_{Py} = 3.4\times10^{-3}$), resulting in a spin-mixing conductance of $g_{\uparrow\downarrow} = 1.95\times10^{19}$ m$^{-2}$, *i.e.*, three times lower than in previous Py/Gr samples. In the second experiment, a 20 nm-thick Cu spacing layer was inserted in between the Py and graphene (*i.e.*, a Py/Cu/Gr stacking), and FMR results compared to those in Py and Py/Cu samples. The rationale is to use the Cu layer as a structural spacer between the Py and the single-layer graphene in order to maintain the Py film unchanged. Note that a thin Cu film does not contribute to the absorption/diffusion of the spin pumped away from the Py film, since Cu has a substantially larger spin diffusion length than the Cu layer thickness used in these experiments[22]. Again, a clear enhancement of the Gilbert damping is observed when graphene is present (*i.e.*, $\Delta\alpha_{Py/Cu/Gr} = \alpha_{Py/Cu/Gr} - \alpha_{Py/Cu} = 4.2\times10^{-3}$), resulting in a spin-mixing conductance of $g_{\uparrow\downarrow} = 2.38\times10^{19}$ m$^{-2}$, i.e., comparable to the values obtained in Gr/Py samples.

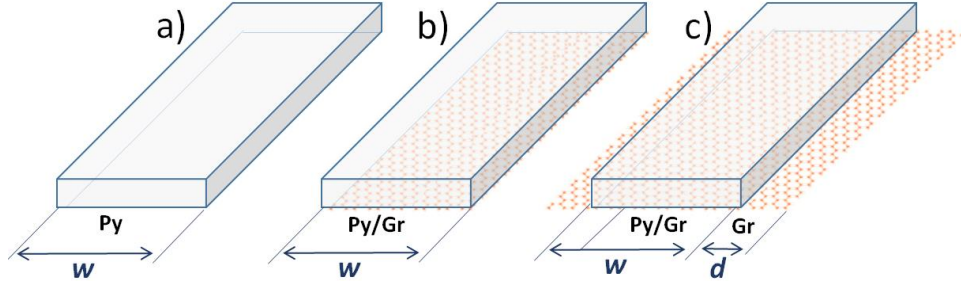

*Figure 1:* *Sketches illustrating the strips used in the experiments. The Py strips are all the same dimensions, with a length of 3 mm and a width w = 25µm. a) Py strip. b) Py/Gr strip. c) Py/Gr-prt strip, with graphene protruding from the sides of the Py strip by d = 20 µm.*

The set of experiments described above eliminate structural changes in the Py as a possible cause for the observed damping enhancement. However, the presence of the ferromagnet in close proximity to the single-layer graphene, even in areas away from the FMR excitation, may influence the diffusion of the spins pumped away from the Py film, which can still act as a spin sink since electrons can flow back into it. To avoid this situation, we patterned the Py/Gr film into long ($l$ = 3mm) and narrow ($w$ = 25 µm) strips that are placed directly underneath the central line of the µ-CPW, as shown in Fig. 1 and in the insets to Fig. 2. Essentially, we prepared three different samples for this study: *a*) a Py strip (Fig. 1a); *b*) a Py/Gr strip (Fig. 1b); and, *c*) a Py/Gr strip with the single-layer graphene protruding away from the Py strip on both sides, which we shall call Py/Gr-prt henceforth (Fig.1c). The upper inset to Fig. 2 shows a scanning electron microscope image of a Py/Gr-prt strip, where one can clearly see the continuous sheet of graphene extending away from the central ferromagnet strip. Note that the length by which graphene protrudes on each side of the ferromagnet strip, *i.e.*, $d$ ~12 µm, is larger than the spin diffusion length of CVD graphene ($\lambda_s$ ~2 µm)[23], in order to allow for a total relaxation of the spin pumped away from the ferromagnet. The devices are prepared by transferring a single-layer graphene onto a GaAs(undoped)/SiO$_2$(100nm) substrate, after which unwanted graphene areas are etched away using a photoresist mask and standard optical lithography. Following etching and e-beam evaporation of the Py strip, a 100 nm-thick layer of silicon oxide is grown atop to insulate the device from the central line of the μ-CPW, which ultimately covers the sample (as depicted in the lower inset to Fig. 2). This geometry guarantees a homogeneous FMR excitation of the whole Py strip.

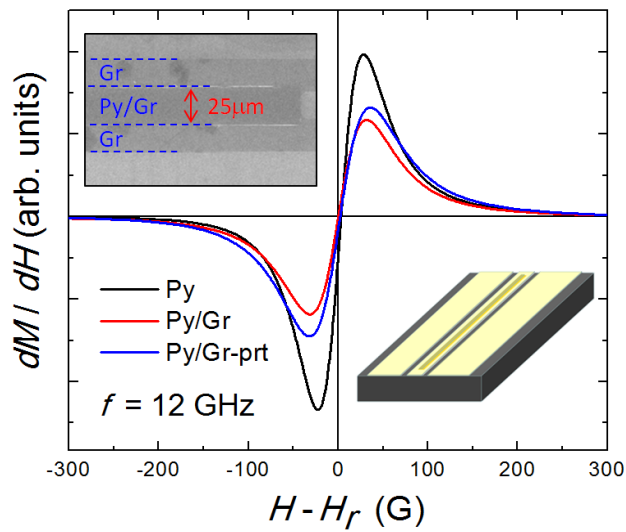

*Figure 2:* *Field-derivative of the FMR response for the three strips measured (Py: Permalloy Gr: graphene and Gr-prt: graphene protruding from the sides of the Py strip). The upper inset shows an electronic microscope image of the Py/Gr-prt stripe before being placed underneath the central line of the μ-CPW sensor (lower inset).*

Standard broadband FMR measurements are performed on the samples described above to extract the FMR linewidths. The corresponding field-derivatives, *dM/dH*, obtained at an irradiation frequency of 12 GHz with the dc magnetic field applied in the plane of the Py strips are shown in Fig. 2. The FMR linewidth, defined as the peak-to-peak distance in the dM/dH data, is the largest for the sample with graphene protruding away from the Py strip (*i.e.*, the Py/Gr-prt strip) and the smallest for the sample with Py only (*i.e.*, the Py strip). The frequency dependence of the in-plane excited FMR linewidth for these samples is shown in Fig. 3a. The observed linear frequency dependence of the linewidth can be explained by means of the dynamical Gilbert damping model, using the following expression:

$$\delta H = \delta H_0 + \frac{4\pi\alpha}{\sqrt{3}\gamma} f \quad , \qquad (1)$$

where $\gamma = g\mu_B / h$ is the gyromagnetic ratio and $\alpha$ is the damping parameter, which is related to the Gilbert damping through the expression $G = \alpha\gamma M_S$, with $M_S$ being the saturation magnetization. The two different contributions to the damping in Eqn. 1 are: a) a sample dependent inhomogeneous part (first term in the equation), which can be calculated from the intercept at zero extrapolated frequency and it does not depend on frequency; and, *b*) dynamical damping (second term in the equation), which scales linearly with frequency and from whose slope the Gilbert damping can be calculated. These extracted damping parameters for in-plane FMR excitation (Fig. 3a) are as follows: $\alpha_{Py} = 9.1\times10^{-3}$ & $G_{Py} = 0.239$ GHz; $\alpha_{Py/Gr} = 11.3\times10^{-3}$ & $G_{Py/Gr} = 0.299$ GHz; $\alpha_{Py/Gr\text{-}prt} = 13.0\times10^{-3}$ & $G_{Py/Gr\text{-}prt} = 0.333$ GHz. There is a considerable enhancement in damping when going from the Py-only strip to the Py/Gr strip, where graphene is only present underneath the ferromagnet. This damping cannot be attributed to spin pumping given the 2D nature of graphene, which is located only underneath the Py strip and does not provide any conduction channel perpendicular to the interface. Indeed, it has been shown that graphene can act as an effective tunnel barrier for electrical spin injection into silicon due to the very large resistivity of carriers across the graphene sheet[24]. Most likely, the observed FMR broadening is due to changes in the magnetic response of the Py due to surface changes induced by the graphene, such as an enhancement of two-magnon scattering processes. It has been recently shown that the deposition of Co films on graphene results in magnetic variations and enhanced magnetic coercivity[16]. In our case, for example, we observe a slight change (<10%) in the magnetization saturation when graphene is present (not shown here). Out-of-plane excited FMR measurements on these strips seem to support this hypothesis. Two-magnon processes are substantially weaker when the precession of the magnetization is excited with the dc field out of the plane, which would explain the similar $\delta H$ vs. $f$ slopes for Py and Py/Gr strips in the out-of-plane excited FMR data of Fig. 3b (black and red data).

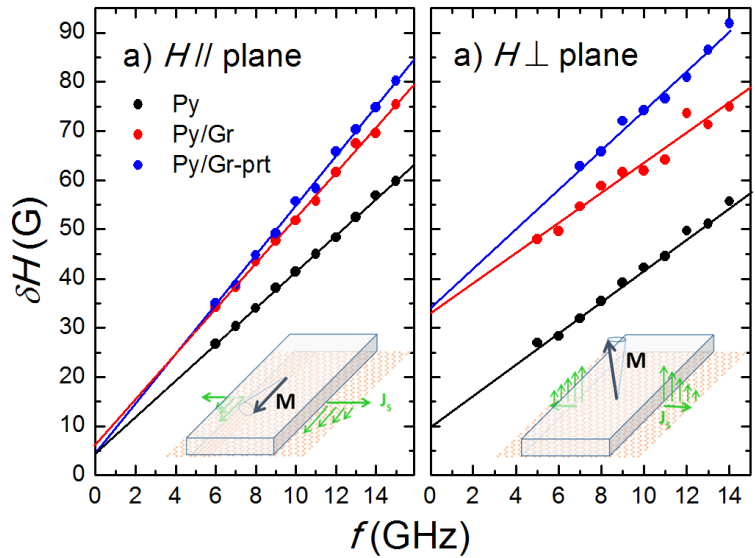

*Figure 3: a) In-plane and b) out-of-plane frequency dependences of the FMR linewidth of the three strips measured (Py: Permalloy Gr: graphene and Gr-prt: graphene protruding from the sides of the Py strip). The insets show the FMR field excitation situations and the corresponding directions of propagation and polarization of the pumped spin currents ($J_s$) into the graphene area protruding away from the edges of the Py strip.*

The central result of this work is the clear enhancement of the dynamical damping observed in Py/Gr-prt strips under both in-plane and out-of-plane excited FMR (blue data in Figs. 3a and 3b). Importantly, this additional damping can only result from the relaxation of spins in the area of graphene away from the Py strip, where an enhanced relaxation due to proximity effects as discussed in the introduction for extended films is not an option. Comparing the FMR broadening in the Py/Gr and Py/Gr-prt strips, and using Eqn. (1) we extract a change in the damping parameter $\Delta \alpha_{Py/Gr\text{-}prt} = \alpha_{Py/Gr\text{-}prt} - \alpha_{Py/Gr} = 1.3 \times 10^{-3}$, resulting in a spin-mixing conductance of $g_{\uparrow\downarrow} = 6.89 \times 10^{18}$ m$^{-2}$. These are a factor 2-3 smaller than the values found in experiments performed on extended films but still comparable to those found in Py/Pt or Py/Pd samples.

We now discuss the fundamental implications of the experiments described above. The observed additional damping enhancement provided by the protruding graphene supports our assertion that spin pumping must occur across the quasi-one-dimensional Py/Gr-prt interface at the very edge of the Py strip. The picture we propose is the following. The proximity of Py to graphene induces a weak equilibrium ferromagnetization in the latter[25]. The precessing magnetization in the Py film pumps a spin current into the graphene layer underneath the Py, thus creating an additional non-equilibrium spin accumulation in that layer. Part of the excess spin polarization is relaxed by local defects and impurities present on graphene (through local-moment scattering or spin-orbit coupling). When the graphene layer does not protrude away from the Py, the remainder non-equilibrium spin accumulation creates a coherent backflow spin current into the Py. Thus, in steady state, there is no net spin current and the enhanced damping of the FMR is mainly due to spin relaxation in the graphene layer underneath the Py. However, when the graphene layer extends beyond the Py, the non-equilibrium spin accumulation causes a spin current to flow into the protruded graphene regions, reducing the amount of coherent backflow into the Py and thus increasing the damping of the FMR due to losses of angular momentum. In this case, it is standard to obtain the spin-mixing conductance associated to the pumping of spin into the extended graphene regions through the expression[26]

$$g_{\uparrow\downarrow} = \frac{4\pi M_s d}{\gamma h} \Delta\alpha_{Py/Gr-prt}. \qquad (2)$$

Yet, this is only justified if the spin current relaxes much faster than the charge diffuses (*i.e.*, when the electronic motion in the extended graphene regions is ballistic or when very strong spin scattering is present, which are unlikely in our samples). We rather expect the charge to diffuse with a relaxation time $\tau \ll \tau_s$, where $\tau_s$ is the spin relaxation time. In this case, a non-equilibrium spin population builds up on the protruding graphene near the Py edge. This causes the spin current to partially diffuse back into the graphene underneath the Py. Thus, the resulting spin-mixing conductance is smaller than that obtained from measuring the excess damping[26]

$$\frac{1}{g_{\uparrow\downarrow}^{actual}} = \frac{1}{g_{\uparrow\downarrow}} - \beta, \qquad (3)$$

where $\beta = 2\tau_s A/(hN_F l \lambda_s)$ is the backflow (dimensionless) parameter, with $N_F$ denoting the density of states of graphene at the Fermi level, $l$ being the length of the Py strip, $A$ being the area of the Py/graphene interface, and $\lambda_s = \sqrt{D\tau_s}$ representing the spin diffusion length (notice that $\lambda_s \ll d$, where $d = 20$ μm is the length of the protruding graphene region). Assuming $\tau_s \approx 10^{-10}$ s, $D \approx 5 \times 10^{-3}$ m²/s, and $N_F = |E_F|/(\pi\hbar^2 v_F^2)$, with $|E_F| \approx 100$ meV and $v_F = 10^6$ m/s (see Ref.[13]) we find $\beta \approx 4 \times 10^{-12}$ m², which is a much larger value (by several orders of magnitude) than the experimental value $1/g_{\uparrow\downarrow} = 1.5 \times 10^{-19}$ m². This implies a *negative* value for $g_{\uparrow\downarrow}^{actual}$ and therefore indicates that Eqn. (3) may not be directly applicable to our setup. We note that this analysis may change in experiments performed in devices using h-BN substrates, for which the spin relaxation parameters would drastically change. H-BN has also been shown to decrease the conductance mismatch in electrical spin injection and therefore may affect spin reflection and relaxation at FM/graphene interfaces [27]. We believe that the main problem in our analysis is not in the estimate of the backflow parameter *β*, since this follows straightforwardly from reasonable estimates for the graphene parameters *l*, *D*, $\tau_s$, and $N_F$. Instead, we believe that the problem lies on the assumption that spin currents pumped by the Py are fully injected into the protruding graphene sheets. It may be possible that very close to the edge where the protruding graphene meets the Py, there is a strongly enhanced spin relaxation, effectively making $\lambda_s$ a much shorter length scale, rendering the backflow negligible. This relaxation could be due to the Py-Gr edge acting approximately as a semiconducting p-n junction, regaining the conductance mismatch that were supposed to be eliminated by the spin pumping method, with graphene underneath the Py being heavily p-doped, while protruding areas are almost undoped. However, this interpretation requires further experimental verification and a more detailed theoretical modeling.

In conclusion, we have presented experimental evidence of a substantial increase in damping in Py/Gr strips when graphene is left to protrude from the sides of the ferromagnet. The surprisingly high spin mixing conductance obtained from the observations raises questions about the physics of dynamical spin injection into two-dimensional structures such as graphene. Our immediate future objective is the direct measurement of the ISHE voltage generated in the protruding graphene region as a result of the pumped spins from the ferromagnet, for which electrodes will be placed at the opposite ends of the protruding graphene lines. However, YIG-based insulating ferromagnetic strips will be used for this purpose, since the low-resistivity Py strip in the present configuration acts as an electrical shunt and prevents the observation of the effect.

**Aknowledgements**

We thank A. Brataas for valuable discussions. S.S. E.M. and E.d.B. acknowledge support from the National Science Foundation (ECCS#1402990). A.A., C.T.C. and B.O. acknowledge support from the Singapore National Research Foundation (Grant Nos. NRFRF2008-07 and NUS-YIA R144-000-283-101).

**Supplemental Material**

Dynamical spin injection at a quasi-one-dimensional ferromagnet-graphene interface

S. Singh[1], A. Ahmadi[1], C.T. Cherian[2,4], E. R. Mucciolo[1], E. del Barco[1], and B. Özyilmaz[2,3,4,5]

[1]Department of Physics, University of Central Florida, Orlando, Florida USA, 32816

[2]Department of Physics, National University of Singapore, 2 Science Drive 3, Singapore 117542

[3]NanoCore, 4 Engineering Drive 3, National University of Singapore, Singapore 117576

[4]Graphene Research Center, National University of Singapore, Singapore 117542

[5]NUS Graduate School for Integrative Sciences and Engineering (NGS), National University of Singapore, Singapore 117456

In this document we present a set of experiments designed to study and prevent changes of the magnetic properties of the Py films due to structural changes associate to the adjacent single-layer graphene and their possible effect in broadening the FMR linewidth of the ferromagnet.

1. **Reversing the FM/Gr stacking order**

We first turn to experiments in which the graphene layer is transferred onto pre-deposited Py films, in contrast with our previous studies[14], where the graphene was transferred onto the substrate (*i.e.*, GaAs) and subsequently Py was deposited atop. Since graphene is not atomically flat when deposited on most substrates[28] (with exception of boron nitride), the growth of the ferromagnet on top of the graphene surface can result in a rough interface and damping contributions due to processes other than the spin pumping, such as two magnon scattering. In addition, the saturation magnetization of the film can also be modified. We have performed nanoscale tunneling microscopy on our graphene films on $SiO_2$ substrates and observed the nanometer sized ripples of our graphene surfaces. To eliminate an enhancement of damping due to structural changes at the interface, we transfer single-layer graphene onto of Py films which have been previously deposited on $SiO_2$ wafers, so that the Py is unaffected, or weakly affected (see Fig. SM1a).

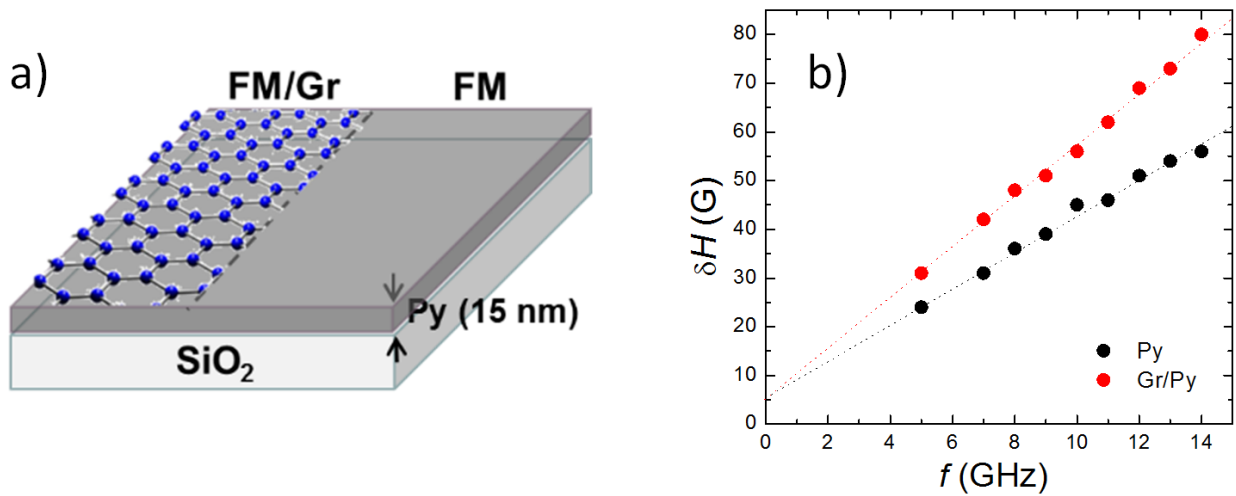

**Figure SM4:** a) Sketch illustrating a Py film in where graphene is wet transferred on the top of the film. b) Frequency dependence of the FMR linewidth of Py and Gr/Py films.

We performed FMR measurements in both Py and Gr/Py by applying the dc magnetic field in the plane of the films. The width of the resonance, in units of external dc field, is extracted by taking the peak-to-peak distance of the FMR field-derivatives. The frequency dependences of FMR linewidth for both samples are shown in figure SM1b. The frequency-dependent width of a FMR resonance represents the effective dynamical damping of the precessing magnetization of the ferromagnet. Any loss of spin angular momentum (in this case due to the presence of graphene) will correspondingly result in an increase in linewidth. Since pure spin currents carry away spin angular momentum, we can theoretically associate the damping enhancement to be a direct consequence of dynamical spin pumping from the Py into the graphene layer.

The damping parameters extracted for these samples are: $\alpha_{Py} = 9.6 \times 10^{-3}$ & $G_{Py} = 0.257$ GHz; $\alpha_{Py/Gr} = 13.0 \times 10^{-3}$ & $G_{Py/Gr} = 0.339$ GHz. This corresponds to a change in the damping parameter ($\Delta\alpha$) of 35%, as compared to 88% change observed in our previous studies of films with the Py deposited onto of the graphene layer[14]. The difference in the relative damping change just by reversing the stacking order of the Py and graphene layers is indicative that a good fraction of the FMR broadening in Py/Gr samples was likely due to structural changes in the surface of the Py film, which has been deposited on top of the rough graphene surface. It is possible that in our experiment the wet transfer of graphene into only one of the Py films causes some degree of contamination (*e.g.*, oxidation of the Py surface) altering the magnetic properties of the ferromagnetic film or the spin polarization at the interface. Recent experiments have shown how contamination during the wet transfer process of two-dimensional crystals can decrease spin polarization at the interface[29]. We do not observe any appreciable change in the saturation magnetization, so such contamination is not drastically changing the magnetostructural properties of the Py film. Nevertheless, the substantial amount of damping observed in the Gr/Py samples of Fig. SM1 indicates that spin pumping at the Gr/Py interface is likely the main cause of FMR broadening.

**2. Inserting a Cu spacing layer in between the Py film and graphene**

It was brought to our attention that while graphene is transferred onto evaporated Py films from an aqueous solution there is a possibility that the Py is oxidized, causing an enhancement of damping of structural origin. Although we find this possibility unlikely to explain the observed results, we made controlled extended films by inserting a thin layer of Copper (Cu) in between the Py film and graphene, and compared the FMR results with Py/Cu films without graphene underneath. The actual samples that we developed are: a) bare Py (14nm); b) Py(14nm)/Cu (20nm); and, Py(14nm)/Cu(20nm)/Gr. The cartoons representing samples are shown in the inset to Fig. SM2a. We use 20nm-thick Cu because it is substantially less than the spin diffusion length and one expects to see negligible spin relaxation and hence no enhanced damping due to the presence of Cu itself in the FMR experiments[22]. In addition, the effect of a graphene layer 20nm away from the Py sample will be negligible and the Py/Cu interface will remain unaltered, whether or not graphene is underneath the heterostructures.

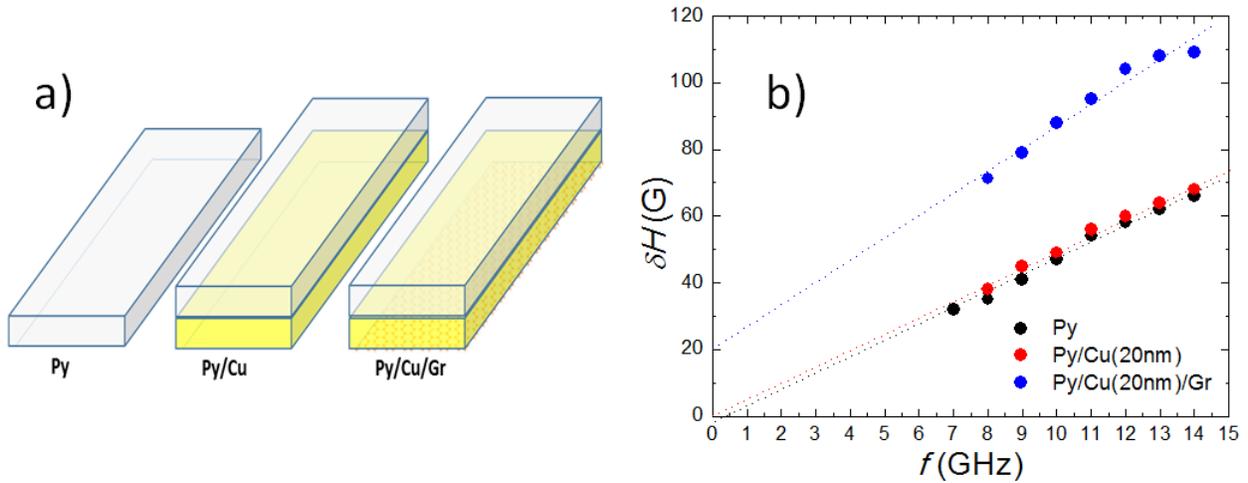

**Figure SM5:** a) Illustrations of the Py, Cu and Gr heterostructures. b) Frequency dependence of the FMR linewidth of Py, Py/Cu and Py/Cu/Gr films.

The in-plane excited FMR linewidths for the above mentioned samples are presented as a function of frequency in Fig. SM2b. As expected, there is no change in damping due to presence of the Cu film underneath the Py, since the pumped spins reflect coherently back into the ferromagnet (no diffusion in the Cu film). One would need to increase the thickness of the Cu film beyond the spin diffusion length for pumping-induced damping to be observable[22]. Now, in the case of the Py/Cu/Gr stacking we do observe a clear change in the damping (i.e. the slope of the frequency dependence of FMR linewidth changes). The extracted damping parameters for these samples are: $\alpha_{Py} = 12\times10^{-3}$ & $G_{Py} = 0.315$ GHz; $\alpha_{Cu/Py} = 11.9\times10^{-3}$ & $G_{Cu/Py} = 0.312$ GHz; and, $\alpha_{Gr/Cu/Py} = 16.2\times10^{-3}$ & $G_{Gr/Cu/Py} = 0.422$ GHz. This is about 35 % change in damping due to the presence of graphene underneath the Cu film, which is similar to what we observe in experiments with graphene transferred atop of the Py films.

### 3. Comparison between different experiments

A first inclination is to attribute the observed damping enhancements to spin pumping into the whole graphene area underneath the ferromagnet. However, one needs to keep in mind that graphene is a two dimensional material with no conduction channels perpendicular to the plane. For standard 3D FM/NM heterostructures where spin pumping has been studied[7], the spin gradient is perpendicular to the FM/NM interface, i.e. the spin polarization decays perpendicularly to the plane. In the case of graphene, the only direction spins can decay is along the plane of the graphene sheet, and this needs to be away from the microwave excitation area. In our above explained experiments on extended thin films only a small part (~20 um × 1000 μm) of the total film (~5 mm × 5 mm) is excited by to the central line of the μ-CPW. Therefore, the spin density pumped from the excited part of the ferromagnet into the graphene directly underneath can only decay in the horizontal graphene plane which is not under microwave excitation. Unfortunately, in experiments with extended films (as those presented in this Supplemental Information document), the areas of graphene away from excitation are in close proximity to the ferromagnet, which is covering the whole film. Given that graphene is highly sensitive to its environment[30], the presence of the ferromagnet is expected to change the intrinsic electronic and spin relaxation properties and result in an enhanced diffusion of the spin currents. This asks for a device geometry in which the damping associated with spin pumping can be systematically differentiated, which is the purpose of the experiments discussed in the main text of this article.

For the sake of clarity, we have compared the change in damping observed for different configuration samples in Table SM1, including the results in Py/Gr-prt strips discussed in the main text, where the enhanced spin diffusion due to the proximity of the ferromagnet in areas of graphene away from the excitation has been eliminated. Although being the smallest, the 11% damping enhancement observed in the latter case can be univocally associated to spin pumping into single-layer graphene and still stands comparable to values observed in heavy metals. It has been shown that a small density of adatoms (*e.g.*, atomic hydrogen) on the graphene surface can induce local deformations resulting in huge enhancements of spin orbit coupling[11]. Recently, a large spin Hall Effect has also been reported for CVD grown graphene (similar to what is used for this work) due to presence of residual Cu adatoms at the graphene surface. Although Cu itself has low spin-orbit interaction, it can induce local deformation in the graphene lattice which results in an enhancement of the spin-orbit interaction.

| Sample | Change in damping Δα | Possible damping contributions |
|---|---|---|
| Py/Gr/Substrate | 88% | Interface roughness damping + enhanced damping due to Py/Gr hybridization + spin pumping |
| Gr/Py/Substrate | 34% | Enhanced damping due to Py/Gr hybridization + spin pumping |
| Py/Cu/Gr/Substrate | 35% | Enhanced damping due to Cu/Gr hybridization + spin pumping |
| Py/Gr/Substrate strips (graphene only under Py and homogeneously irradiated) | 24% | Interface roughness damping |
| Py/Gr/Substrate strips (graphene protruding out of Py) | 15% | Spin pumping only |

**Table SM1:** Comparison between damping enhancements for all the samples discussed in this article.